\documentclass{PHYEAUTH}
\usepackage{graphicx}
\usepackage{amsmath}
\usepackage{amssymb}

\def\be{\begin{equation}}
\def\ee{\end{equation}}
\def\bea{\begin{eqnarray}}
\def\eea{\end{eqnarray}}
\def\ve{\varepsilon}
\def\wc{\omega_c}
\def\ve{\varepsilon}
\def\w{\omega}

\def\tq{\tau_{\rm q}}
\def\tin{\tau_{\rm in}}
\def\ttr{\tau_{\rm tr}}
\begin{document}

\begin{frontmatter}

\title{Fractional microwave-induced resistance oscillations}

\author[address1,address3]{I.A. Dmitriev\thanksref{thank1}},
\author[address1,address2,address4]{A.D. Mirlin}
and
\author[address1]{D.G. Polyakov}

\address[address1]{Institut f\"ur Nanotechnologie, Forschungszentrum
Karlsruhe, 76021 Karlsruhe, Germany}

\address[address2]{Institut f\"ur Theorie der kondensierten Materie,
Universit\"at Karlsruhe, 76128 Karlsruhe, Germany}

\address[address3]{Also at A.F.~Ioffe Physico-Technical Institute,
194021 St.~Petersburg, Russia.}

\address[address4]{Also at Petersburg Nuclear Physics
Institute, 188350 St.~Petersburg, Russia.}

\thanks[thank1]{
Corresponding author.
E-mail: dmitriev@int.fzk.de}

\begin{abstract}
We develop a systematic theory of  
microwave-induced oscillations in magnetoresistivity of a 2D electron gas
in the vicinity of fractional harmonics of the cyclotron resonance, observed in
recent experiments.
We show that in the limit of well-separated Landau levels the effect is
dominated by the multiphoton inelastic mechanism. At moderate magnetic
field, two single-photon mechanisms become important.
One of them is due to resonant series of multiple single-photon
transitions, while the other originates from microwave-induced
sidebands in the density of states of disorder-broadened Landau levels.
\end{abstract}

\end{frontmatter}

\section{Introduction}\noindent
Recently, much attention has been attracted to the discovery of
microwave-induced resistance oscillations (MIRO) \cite{zudov01}, followed by the
spectacular observation of zero-resistance states (ZRS) in the oscillation
minima \cite{mani02,zudov03}.
Two microscopic mechanisms of the MIRO have been proposed: the ``displacement''
mechanism
related to the effect of microwaves on the impurity scattering
\cite{ryzhii,durst03,VA},  and the ``inelastic'' mechanism accounting for
nonequilibrium oscillatory changes in the electron distribution
\cite{dmitriev03,long}. 
Both mechanisms rely on the energy oscillations of the density of states (DOS)
$\nu(\ve)$ of disorder-broadened Landau levels (LLs) and reproduce the observed
phase of the $\w/\wc$--oscillations (here, $\w$ and $\wc=eB/mc$ are the
microwave and the cyclotron frequencies). The inelastic mechanism yields
temperature-dependent MIRO with the amplitude, proportional to the inelastic
scattering time $\tau_{\rm in}\propto T^{-2}$, while the displacement
contribution is $T$-independent, in disagreement with the experiments. At
relevant $T\sim 1$~K, the inelastic effect dominates and the corresponding
theory \cite{dmitriev03,long} reproduces the experimental
observations \cite{zudov01,mani02,zudov03}.

Further experimental investigations at elevated microwave power led to
the discovery
of ``fractional'' MIRO and ZRS \cite{multi,dorozhkin06} located near
the fractional
harmonics of the cyclotron resonance, $\w/\wc=1/2,\,3/2,\,5/2,\,2/3..$, to be 
contrasted with the integer MIRO \cite{zudov01,mani02,zudov03}. These remarkable
observations motivated the present study where we
address multiphoton effects 
and effects of the microwave radiation on the electronic spectrum, which govern
the fractional MIRO in the case of separated LLs.

\section{Inelastic mechanism of the MIRO}\noindent
We start by including the multiphoton processes in the
theory \cite{dmitriev03,long}.  
In a classically strong magnetic field, $\wc\ttr\gg1$, the diagonal resistivity
$\rho_{xx}$ reads \cite{dmitriev03}
\be\label{QDrude}
\rho_{xx}/\rho_{xx}^D=\int\!d\ve \tilde{\nu}^2(\ve)\partial_\ve f(\ve),\qquad 
\ee
where $\rho_{xx}^D=(e^2v_F^2\nu_0\ttr)^{-1}$ is the Drude resistivity,
$v_F$ the Fermi velocity, $\ttr$ the transport scattering time, $\nu_0=m/2\pi$,
and $\tilde{\nu}(\ve)=\nu(\ve)/\nu_0$ the dimensionless density of states (DOS)
of disorder-broadened LLs. The MIRO originate from the
microvave-induced oscillations in the distribution function $f(\ve)$,
which obeys the kinetic equation
\be\label{kineq}
f(\ve)\!-\!f_T(\ve)\!=\!\frac{\tin}{4\tq}\sum_n \!A_n
\tilde{\nu}(\ve\!-\!n\w)[f(\ve\!-\!n\w)\!-\!f(\ve)].
\ee
Here $A_n=A_{-n}$ describes the probability of $n$-photon absorption
(emission) and $f_T(\ve)$ is the thermal distribution. The leading contribution
to the integer MIRO comes from the single-photon $A_1=A_{-1}\equiv
P_\w$, where
\be\label{A1}
P_\w=(\tq/\ttr)(ev_F E_\w)^2/\w^2(\w+\wc)^2\,.
\ee
Here $E_\w$ is the amplitude of
the circularly polarized
microwave field \cite{polarization}, and $\tq\ll\tin,\ttr$ is the total
(quantum) disorder-induced
scattering time. If one assumes, in accord with the experimental conditions,
that the temperature is high, $2\pi^2 T/\wc\gg1$, the contribution to
$\rho_{xx}$
of first order in $P_\w$ takes the form \cite{dmitriev03,long}
\be\label{1order}
\rho_{xx}/\rho_{xx}^D=\langle\tilde{\nu}^2(\ve)\rangle_\ve+(\tin/4\tq)P_\w
F(\w),
\ee
where $\langle\ldots\rangle_\ve$ denotes the averaging over the period $\wc$ of
the DOS, and the function $F(\Omega)$ oscillates with $\Omega/\wc$,
\[
F(\Omega)=\Omega\,\langle\tilde{\nu}^2(\ve)\,
\partial_\ve[\,\tilde{\nu}(\ve+\Omega)+\tilde{\nu}(\ve-\Omega)\,]\,\rangle_\ve.
\]

\begin{figure}[h]
\begin{center}\leavevmode
\includegraphics[width=0.7\linewidth]{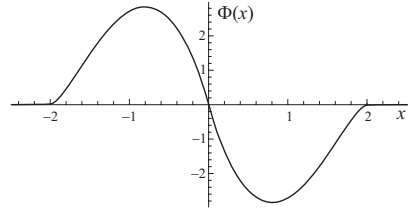}
\caption{Function $\Phi(x)$, Eq.~(\ref{phi}).} \label{fig1}
\end{center}
\end{figure}

In the limit of separated LLs, $\wc\tau_{\rm q}\gg 1$, the DOS
$\tilde{\nu}(\ve)=\tau_{\rm q}{\rm Re}\sqrt{\Gamma^2-(\delta\ve)^2}$ is a
sequence
of semicircles of width $2\Gamma=2(2\wc/\pi\tau_{\rm q})^{1/2}$, where
$\delta\ve$ is the detuning from the center $(n\!+\!\frac{1}{2})\wc$ of the
nearest
LL. In that case
\be\label{F}
F(\Omega)=(16\Omega\wc^2/3\pi^2\Gamma^3)\,\Phi[(\Omega-N_\Omega\wc)/\Gamma],
\ee
where $N_\Omega$ is the integer number closest to $\Omega/\wc$, and the odd
function
$\Phi(x)$ is nonzero at $|x|<2$ (Fig.~\ref{fig1}) 
\be\label{phi}\Phi(x)\!=\! x(1+|x|)\sqrt{|x|(2-|x|)}-3x\,{\rm
arccos}(|x|-1)\,.\ee
  \begin{figure}[h]
\begin{center}\leavevmode
\includegraphics[width=0.7\linewidth]{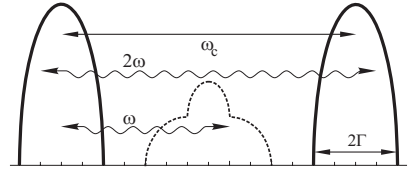}
\caption{Illustration of the processes leading to the fractional MIRO,
for $\w/\wc=1/2+\Gamma/2$ and $\wc/\Gamma=7$. Single-photon transitions within
and between LLs (solid lines) are forbidden, while two-photon processes are
allowed. The microwave-induced sidebands (\ref{sb}) (dashed line) make
single-photon processes possible.} \label{fig2}
\end{center}
\end{figure}
\section{Inelastic multiphoton mechanism of the fractional MIRO}\noindent
Provided $\Gamma\ll\wc$, the oscillatory part of $\rho_{xx}$ (\ref{1order}) is
finite only in narrow intervals $(\w-N\wc)<4\Gamma$ around the integer
values $\w/\wc=N$ \cite{windows}. Outside these intervals
$\nu(\ve)\nu(\ve+\w)\equiv0$. Therefore, single-photon absorption is forbidden,
so that $f(\ve)=f_T(\ve)$ as long as multiphoton processes, given by $A_n$ with
$|n|>1$, are not taken into account. Inclusion of two-photon processes leads
to the appearance of the fractional MIRO in the frequency intervals
$|\w-(N+1/2)\wc|<\Gamma$, $N=0,1,2..$, where $\nu(\ve)\nu(\ve+2\w)\neq0$ (see
Fig.~\ref{fig2}). In these intervals,
\be\label{2photon}
\rho_{xx}/\rho_{xx}^D=\langle\tilde{\nu}
^2(\ve)\rangle_\ve+(3\tin/32\tq)P_\w^2 F(2\w),
\ee
where we used $A_2=A_{-2}=3P_\w^2/8$ \cite{VA,mechanisms}.
The doubling of the argument of the function $F$ in
Eq.~(\ref{2photon}) [as compared to the integer MIRO,
Eq.~(\ref{1order})] reflects the two-photon nature of the effect
and leads to the emergence of the fractional MIRO at half-integer $\w/\wc$. The
form and phase of the fractional oscillations (\ref{2photon})
reproduce those for the integer MIRO, Eq.~(\ref{1order}). 
Similarly to the integer case, there exists \cite{leiliumulti} a multiphoton
contribution to the fractional MIRO governed by the displacement mechanism
\cite{ryzhii,durst03,VA}, which has a similar form but is
a factor $\wc\tin/\tq\Gamma\gg 1$ smaller
\cite{unpublished} than the inelastic one (\ref{2photon}).

  With increasing microwave power $P_\w$, the resistivity
(\ref{2photon}) in the oscillation minima becomes negative, which indicates
a transition to the ZRS \cite{andreev03}. Remarkably, like in the integer case
\cite{long}, the leading-order approximation (\ref{2photon}) for the
multiphoton inelastic effect is sufficient to describe the
fractional photoresponse even at such high power, since the second order
contribution $\propto (P_\w^2\tin/\tq)^2$ remains small in the parameter
$\Gamma/\wc$.
\section{Sideband mechanism of the fractional MIRO}\noindent
Using the formalism developed in \cite{VA}, it can be shown that the microwave
illumination results in the appearance of "sidebands" in the DOS, located at
distance $\w$ on both sides of every LL (see Fig.~\ref{fig2}). To first order
in $P_\w$ and assuming again $|\w-(N+1/2)\wc|<\Gamma$, we obtain the following
expression for the microwave-induced sidebands \cite{unpublished}:
\be\label{sb}
\tilde{\nu}^{(sb)}(\ve)=(\pi
P_\w/8\wc\tq)\,[\,\tilde{\nu}(\ve+\w)+\tilde{\nu}(\ve-\w)\,]~,
\ee
where $\tilde{\nu}(\ve)$ is the unperturbed DOS.
In the presence of the sidebands, single-photon transitions become
possible (Fig.~\ref{fig2}),
$\tilde{\nu}(\ve)\tilde{\nu}^{(sb)}(\ve\pm\w)\neq0$, resulting in
the ``sideband'' contribution to the fractional MIRO,
\be\label{rhosb} \rho_{xx}^{(sb)}/\rho_{xx}^D=(\pi\tin/64\wc\tq^2)
\,P_\w^2\,F(2\w)\,, \ee
which has the same form as  the leading two-photon
inelastic contribution (\ref{2photon}), but is smaller a factor
$\wc\tq$. One more contribution originates from the sidebands oscillating in
time with frequency $2\w$. ``Oscillating sideband'' contribution
\cite{unpublished} is symmetric with respect to the detuning from the fractional
resonances, in contrast to the antisymmetric $F(2\w)$, and is
a factor $(\wc\tq)^{1/2}$ smaller than the two-photon contribution
(\ref{2photon}).

\section{Conclusion}\noindent
In the limit of well-separated LLs, $\wc\gg\Gamma$, the
fractional MIRO are governed by the multiphoton inelastic mechanism,
Eq.~(\ref{2photon}). At $\wc\sim\Gamma$, the sideband contribution (\ref{rhosb})
becomes relevant. Close
to $\wc$ at which LLs start to overlap; specifically, 
at $\wc<4\Gamma$,
the
effect is dominated by the resonant
series of multiple single-photon transitions
\cite{dorozhkin06,crossover}. This effect appears at order
$(\tin P_\w/\tq)^2$. In the limit of strongly overlapping LLs, the
fractional features get exponentially suppressed with respect to the integer
MIRO \cite{crossover,mechanisms}.

We thank S.I.Dorozhkin and
M.A.~Zudov for information about the experiments,
 and I.V.~Gornyi for stimulating
discussions. This work was
supported by the SPP ``Quanten-Hall-Systeme'' 
and Center for Functional Nanostructures 
of the DFG, by INTAS Grant
No.~05-1000008-8044, and by the RFBR.


\begin{thebibliography}{8}
\bibitem{zudov01} M.A.~Zudov, R.R.~Du, J.A.~Simmons, and J.R.~Reno, Phys.\ Rev.\
B 64 (2001) 201311(R).
\bibitem{mani02} R.G.~Mani, J.H.~Smet, K.~von~Klitzing, V.~Narayanamurti,
W.B.~Johnson, and V.~Umansky, Nature 420 (2002) 646.
\bibitem{zudov03} M.A.~Zudov, R.R.~Du, L.N.~Pfeiffer, and K.W.~West, Phys.\
Rev.\ Lett.\ 90 (2003) 046807.
\bibitem{ryzhii} V.I.~Ryzhii, Sov.\ Phys.\ Solid State 11 (1970) 2078.
\bibitem{durst03} A.C.~Durst, S.~Sachdev, N.~Read, and S.M.~Girvin, Phys.\ Rev.\
Lett.\ 91 (2003) 086803.
\bibitem{VA} M.G.~Vavilov and I.L.~Aleiner, Phys.\ Rev.\ B 69 (2004) 035303.
\bibitem{dmitriev03} I.A.~Dmitriev, A.D.~Mirlin, and D.G.~Polyakov,  Phys.\
Rev.\ Lett.\ 91 (2003) 226802.
\bibitem{long} I.A.~Dmitriev, M.G.~Vavilov, I.L.~Aleiner, A.D.~Mirlin, and
D.G.~Polyakov, Phys.\ Rev.\ B 71 (2005) 115316.
\bibitem{multi} M.A.~Zudov, R.R.~Du, L.N.~Pfeiffer, and K.W.~West, Phys.\ Rev.\
B 73 (2006) 041303(R).
\bibitem{dorozhkin06} S.I.~Dorozhkin, J.H.~Smet, K.~von~Klitzing, L.N.~Pfeiffer,
and K.W.~West, cond-mat/0608633.
\bibitem{polarization}Here we consider passive circular polarization
of the microwave field. Qualitatively similar results for arbitrary
polarization will be presented elsewhere \cite{unpublished}.
\bibitem{windows} S.I.~Dorozhkin, J.H.~Smet, V.~Umansky,
and K.~von~Klitzing, Phys.\ Rev.\ B 71 (2005) 201306(R).
\bibitem{mechanisms} I.A.~Dmitriev, A.D.~Mirlin, and D.G.~Polyakov, Phys. Rev. B
75 (2007) 245320.
\bibitem{leiliumulti} X.L.~Lei and S.Y.~Liu, Appl.\ Phys.\ Lett.\ 88 (2006)
212109.
\bibitem{andreev03} A.V.~Andreev, I.L.~Aleiner, and A.J.~Millis, Phys.\ Rev.\
Lett.\ 91 (2003) 056803.
\bibitem{unpublished} I.A.~Dmitriev, A.D.~Mirlin,
and D.G.~Polyakov, Phys. Rev. Lett. 99 (2007) 206805.
\bibitem{crossover} I.V.~Pechenezhskii, S.I.~Dorozhkin, and I.A.~Dmitriev, JETP
Lett.\ 85 (2007) 86.

\end{thebibliography}
\end{document}